# Indirect Job-Shop coding using rank: application to QAOA (IQAOA)

Eric Bourreau[a], Gérard Fleury[b] and Philippe Lacomme[b]

[a] Université de Montpellier, LIRMM, 161 Rue Ada, 34095 Montpellier, France - eric.bourreau@lirmm.fr
[b] Université Clermont Auvergne, Clermont Auvergne INP, UMR 6158 LIMOS, 1 rue de la Chebarde, Aubière, 63178, France – gerard.fleury@isima.fr and philippe.lacomme@isima.fr

## 1. Introduction

A job-shop scheduling problem can be formulated as processing a set of $n$ jobs $J = \{J_1, \ldots, J_n\}$ on a set of $m$ machines $M = \{M_1, \ldots, M_m\}$ Each job $J_i$ is an ordered of $m$ operations denoted $O_{i1}, O_{i2}, \ldots, O_{im}$ for which precedence constraints are defined. The $k^{th}$ operation of job $i$ i.e., $O_{ik}$) will be processed using a predefined machine $m_{ik}$ with duration $p_{ik}$. Each machine can process (without pre-emption) only one operation.

The goal is to schedule each operation on the machines, considering precedence constraints, to minimize the total makespan. According to the $\alpha|\beta|\gamma$ notation introduced by Garey and Johnson in 1976, the problem can be denoted as $j||C_{max}$, and it has been proved to be NP-hard (Garey and Johnson, 1976). JSSP is widely acknowledged as one of the most challenging scheduling problems over the last few decades. A useful representation of the problem is the disjunctive graph model proposed by Roy and Sussmann in 1964 (Roy and Sussmann, 1964).

Thanks to the disjunctive graph model, any instance of the job shop problem can be visually represented through a directed graph $G = (V, A, E)$, where $V$ denotes the set of nodes, $A$ represents the set of conjunctive arcs, and $E$ is the set of disjunctive arcs. The set of nodes encompasses an element for each operation $O_{ij}$, a source node denoted by 0 connected to the initial operation of each job, and a sink node denoted by * linked with the final operation of each job.

Conjunctive arcs depict the scheduling of different operations within the jobs, connecting each pair of consecutive operations in the same job. Meanwhile, each pair of disjunctive arcs links two operations, belonging to distinct jobs, scheduled to the same machine.

A feasible solution is represented by an acyclic subgraph that includes all conjunctive arcs and one disjunctive arc for each pair in the set of disjunctive arcs. An optimal solution is modelized by a oriented disjunctive graph with minimal makespan. The job-shop scheduling problem relies on the following assumptions (see for example (Chassaing et al., 2014)):

- Each job comprises a finite number of operations.
- The processing time for each operation on a specific machine is predefined.
- A specific sequence of operations must be followed to complete each job.
- Each job is executed on each machine only once.
- Each machine can process only one job simultaneously.
- The process of one operation cannot be split.
- Preemption is not allowed.
- No due date constraints are specified.
- There are no setup or tardiness costs associated.

A solution is fully defined by
- the definition of the earliest starting times $st_{i,j}$ that meet the Job-shop constraints;
- the definition of the same sequence of jobs on all the machines.

The optimal solution corresponds to an oriented disjunctive graph with the minimal makespan $C_{max}$.

## 2. Indirect Job-Shop Coding using rank

### 2.1. Graph modeling

The principles of Quantum Annealing Optimization (QA) involve a gradual reduction of quantum fluctuations to overcome barriers in energy landscapes while seeking the global minima of a function. Quantum annealers face limitations in resolving optimization problems due to the size of the Hamiltonian circuit, particularly when constraints are incorporated into the objective function using penalty terms. Approaches based on Driven Hamiltonians are devised to circumvent these penalty terms and the requirement for all-to-all connectivity, instead defining new Hamiltonians that commute with the constraints.

Mixer Hamiltonians are calibrated to prioritize the exploration of operational research scenarios where the feasible subspace is significantly smaller than the full space (represented by the set of n-bit strings used for $H$), achieved through a mapping from initial states to final solutions. Focusing solely on the feasible subspace has been a subject of interest within the Operations Research (O.R.) community for decades, and these approaches have proven successful across various research domains, resulting in highly efficient metaheuristics.

In Job-Shop Scheduling, a commonly used indirect representation is based on the Bierwith vector (Bierwith, 1995), which can be efficiently transformed (in $O(n)$ time complexity) into an oriented disjunctive graph representing a solution for the job-shop problem. Additionally, scheduling problems relying on disjunctive graphs exist, take advantages of efficient local search methods based on operator applied to the longest path (as seen in Grabowski's block definition, for instance) (Grabowski et al., 1986).

In flow-based scheduling problems such as RCPSP, local search strategies are centered around cuts (Artigues, 2003), while in routing problems, they often consider the geometric aspects of solutions. Diversification mechanisms vary depending on the metaheuristic used; for example, in simulated annealing (SA), they are temperature-based, whereas in memetic algorithms (MA) they involve mutation (Moscato, 1999). These mechanisms allow for transitions from one solution to potentially worse solutions with a non-zero probability, facilitating the traversal of barriers within the objective function landscape.

In Job-Shop Scheduling, a prevalent indirect representation relies on the Bierwith vector (Bierwith, 1995), which can be efficiently transformed (in O(n) time complexity) into an oriented disjunctive graph representing a single solution for the job-shop. Additionally, scheduling problems based on disjunctive graphs are notable, where local search approaches capitalize on the longest path (as exemplified by Grabowski's block definition, for instance) (Grabowski et al., 1986). Such modelization has been used (Chassaing et al., 2014) (Caumond et al., 2008),

In flow-based scheduling problems like RCPSP, local search strategies center around cuts (Artigues, 2003), while in routing problems, they often leverage the geometric properties of solutions. Diversification mechanisms vary depending on the metaheuristic employed; for instance, they may be temperature-based in simulated annealing (SA) or involve mutation in memetic algorithms (MA) (Moscato, 1999). These mechanisms allow for transitions from one solution to potentially worse ones with a non-zero probability, aiding in overcoming barriers within the objective function landscape.

A well-known representation of solutions for the Vehicle Routing Problem is the giant trip (ordered vector of all nodes to service) that can be mapped into one VRP solution using the Split Algorithm (Lacomme et al., 2001). The Split algorithm defines a mapping from the set of giant trip of a TSP (Traveling Salesman Problem) to solutions of the VRP and a metaheuristic based approach is only used to investigate efficiently the set of giant trips.

In the context of Vehicle Routing Problems (VRP), a widely recognized representation is the "giant trip," which can be converted into a VRP solution through the Split Algorithm (Lacomme et al., 2001). The Split algorithm establishes a mapping from the set of giant trips in a Traveling Salesman Problem (TSP) to VRP solutions, enabling metaheuristic-based approaches to efficiently manipulate the set of giant trips exclusively.

In 1996, Cheng et al. were the first in conducting a comprehensive analysis of the string coding approach and decoding mechanisms within the broader context of optimization. They emphasized that the mapping functions typically exhibit an $m$-to-$n$ type, and for a large part of optimization problem only $n$-to-$1$ mapping function exist meaning that multiple objects in the coding space can represent the same solution (Figure 1).

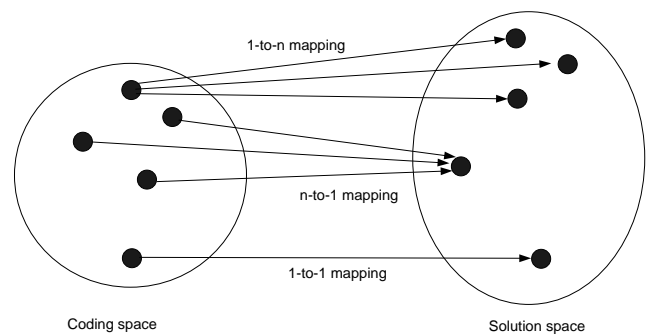

**Fig. 1.** Mapping from coding to solution space chart of a second solution (Cheng et al., 2019)

Their observations span both scheduling and routing, emphasizing the principle of adopting a mapping perspective from the constraints of the problem. This approach aims to establish indirect representations that facilitate the generation of feasible solutions during both the initial phase and the evolutionary process.

The split algorithms, as described in Lacomme et al. (2001), represent a classical approach to routing. They involve transforming a large TSP trip into, for instance, a solution for the VRP, as illustrated in Figure 2.

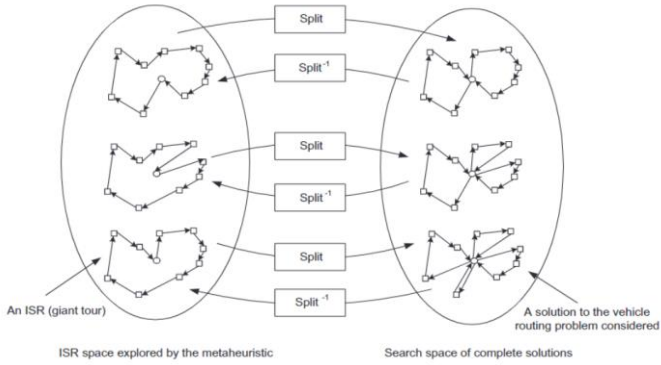

**Fig. 2.** The TSP search space projection into the VRP search space (Lacomme et al., 2001)

In this context, most order-first/split-second methods operate by initially generating an indirect solution representation, often termed a giant tour or task ordering, in the first phase. Feasible routes are then derived from this indirect representation in a subsequent phase. As stressed by (Bourreau et al., 2022), this approach offers several advantages:
- Every solution to the routing problem can be represented by an Indirect String Representation (ISR).
- Through a mapping function, each ISR can be mapped into a solution for the original problem, and this mapping can be performed optimally in very low complexity algorithm.
- There exists at least one indirect representation that can be mapped into the optimal solution of the problem.

It has been demonstrated by (Lacomme et al., 2001) that the local search operator must encompass modifications to a VRP (Vehicle Routing Problem) solution, including basic moves such as 2-OPT and insertion, as well as moves applied to the giant tour. The overall effectiveness of the local search relies on alternating between these two spaces.

Mixer Hamiltonians align with current research trends, aiming to enable the transformation of qubit strings into solutions by meticulously defining the mapping function. This facilitates exploration solely within the "feasible" search space. In contrast to classical operations research fields, the Mixer Hamiltonian inherently embeds explicit local search dynamics. Studies, such as that by Hadfield et al. (2019), have demonstrated the efficacy of this Hamiltonian as a novel and efficient model for various optimization problems.

*2.2. Graph modeling*

Let's examine an example of a Job-Shop comprising 3 jobs and 3 operations, as outlined in Table 1. In this table, each job is associated with a set of operations, and for each operation, the corresponding machine ($M_1$, $M_2$, or $M_3$) and its processing time on that machine are provided. The sequence of operations remains consistent across machines (i.e., it is not dependent on the job), albeit with varying processing times. For instance, Job 1 necessitates 10 units of time on Machine M1, while Job 2 requires 15 units of time on the same machine.

Table 1
Example of JSSP instance data

| Jobs | Operation 1 | Operation 2 | Operation 3 |
|---|---|---|---|
| $i = 1$ | (M1, 10) | (M2, 35) | (M3, 25) |
| $i = 2$ | (M2, 15) | (M1, 6) | (M3, 12) |
| $i = 3$ | (M3, 100) | (M2, 1) | (M1, 10) |

The disjunctive graph illustrating this problem is introduced in Fig.3. In this graph, an arc (in full line) between two successive operations ($O_{i,j}; O_{i,j+1}$) represents the sequence of operation of the job $i$. It is weighted by the minimal time lag between the starting times of these two consecutive operations meaning that $st_{i,j+1} \geq st_{i,j} + p_{i,j}$.

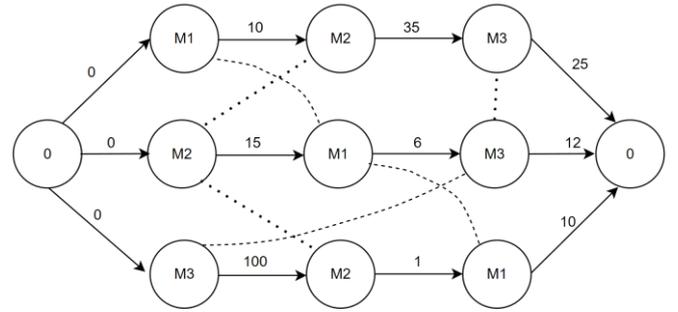

**Fig. 3.** Non-oriented disjunctive graph

Each pair of disjunctive arcs is depicted by a dashed line and symbolizes the resource constraint between two operations that are scheduled to use the same machine.

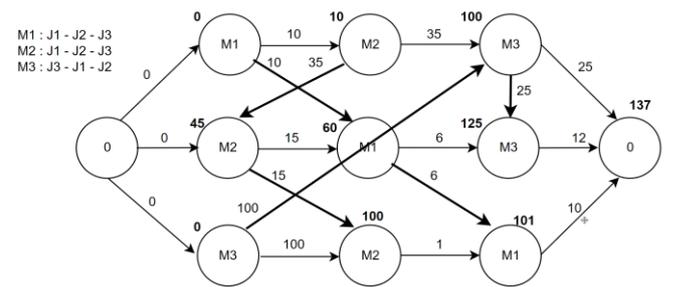

**Fig. 4.** Oriented disjunctive graph

One acyclic conjunctive graph is given in Fig. 4 where all operations in disjunction (one set per machine) are reduced to disjunctives arcs modeling the sequencing of operations processed on the same machine.

The left-shifted solution (semi-active) is obtained by execution of one longest-path algorithm that permits to compute the earliest starting times of operation $st_{i,j}$. The starting time of operation ∗ is 137 and the full solution is introduced in Fig 5.

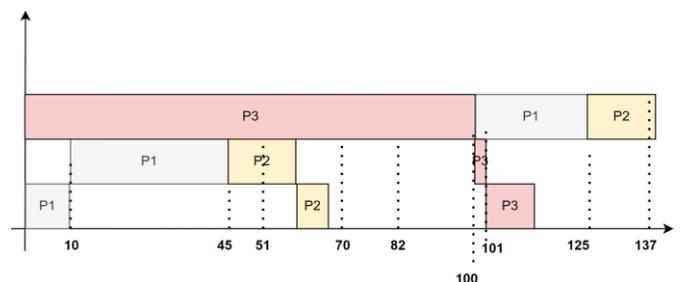

**Fig. 5** Gantt chart of one solution

The oriented disjunctive graph in Fig. 4 is related to the relative order J1, J2, and J3, modeling a solution with a makespan of 137, as presented in Fig. 5. It appears evident that each order of the jobs defines a different oriented disjunctive graph and, therefore, corresponds to a distinct solution.

### 2.3. Quasi-Direct representation

A solution can be derived from the complete job order, providing a quasi-direct representation of the solution. For instance, the solution depicted in Fig. 6 is derived from sequencing Job 1 first, followed by Job 2, and finally Job 3. The significance of defining an ad hoc quasi-direct representation of a solution has been underscored for years. Notably, in a subsequent publication by Cheng et al. (1996), the authors clearly articulated that a quasi-direct representation allows for: (1) a coding space and (2) a solution space. According to Cheng et al. (1996), an effective mapping function must assign a solution to any object within the coding space, ensuring that it satisfies the feasibility constraints.

Regardless of the specific mapping function employed, a job-shop solution manifests as a fully oriented disjunctive graph. Utilizing a longest path computation, one can determine the earliest starting time for operations, thereby defining a semi-active solution.

(Schmid et al., 2024) introduced in 2024 a very similar approach but they do not refer the famous publication of (Bierwith, 1995) who introduce very early the mapping from Bierwith's vector (vector with repetition) and acyclic disjunctive graph modeling solution.

### 2.4. Indirect representation of solutions

The indirect representation of a solution can harness the inherent one-to-one relationship between permutations and a concept known as subexceedant functions, ultimately simplifying the modeling of permutation ranks using a single integer. Various methods exist for establishing this one-to-one correspondence, with the most renowned being the Lehmer code, also referred to as the inversion table. An algorithmic description of this approach is initially presented in Knuth's work from 1981 (Knuth, 1981).

The Bierwith vector are permutation with repetition and for a problem $n$ jobs with $m$ machines assuming that job are processed only one on each machine, the total number of permutation with repetition is:

$$\frac{(n.m)!}{\prod_{i=1}^{n} m!}$$

A algorithm description of both computation of $f$ from one permutation $\sigma$ and computation of a permutation $\sigma$ from $f$ has been provided lately in (Bourreau et al., 2023).

The total number of disjunctive graph is $m!^n$ and represents the total number disjunctive oriented graph . For a 10 jobs, 10 machines Job-Shop instance we have about $3.95.10^{65}$ disjunctive graphs and about $2.3.10^{62}$ Bierwith vectors. Each Bierwith vectors can be mapped to one oriented disjunctive graph modeling a solution.

```
Algorithm 1. Compute_r ()
Input parameters:
    σ : a permutation of n element
    [n] : the interval
Output parameters:
    r : the rang
Local parameters:
    E : number of operation per machine
Begin
1.  E = E = [m, m, …, m]
2.  For i = n − 1 to 1 do
3.      val = σ[i]
4.      right = right + 1
5.      For j = 0 to val − 1 do
6.          If (E[j] > 0) then
7.              remaining = n × m − right
8.              num = remaining!
9.              den = 1
10.             For p = 0 to n − 1 do
11.                 If (p == j) and (E[p]>0) Then
12.                     den = den × (E[p] − 1)!
13.                 Else
14.                     If (E[p]>0) Then
15.                         den = den × E[p]!
16.                     End If
17.                 End If
18.             End For
19.         End If
20.         End If
21.         r = r + num /den
22.     End For
23.     E[p] = E[p] − 1
24. End For
25. Return r
26.End
```
Algorithm 1. Conversion of $\sigma$ into a rank

For a 3 jobs – 3 machines instance, we have about 1680 Bierwith's vectors and each vector is mapped into one graph and each graph is related to the finishing time of the last operation on the last machine. For the instance of Table 1. The list of Bierwith vector is the following list:

```
   0  :  [0, 0, 0, 1, 1, 1, 2, 2, 2]    193
   1  :  [0, 0, 0, 1, 1, 2, 1, 2, 2]    182
   2  :  [0, 0, 0, 1, 1, 2, 2, 1, 2]    182
   3  :  [0, 0, 0, 1, 1, 2, 2, 2, 1]    182
   4  :  [0, 0, 0, 1, 2, 1, 1, 2, 2]    182
   5  :  [0, 0, 0, 1, 2, 1, 2, 1, 2]    182
   6  :  [0, 0, 0, 1, 2, 1, 2, 2, 1]    182
  …
1677  :  [2, 2, 2, 1, 1, 0, 0, 1, 0]    192
1678  :  [2, 2, 2, 1, 1, 0, 1, 0, 0]    192
1679  :  [2, 2, 2, 1, 1, 1, 0, 0, 0]    192
```

The **algorithm 1** consists in computing the rank $r$ related $\sigma$. The algorithm is composed of two loop: the first one (from step 2 to step 24) iterates on $\sigma$ and the second (from step 5 to 22) computes the shift from previous partial evaluated rank.

When $\sigma = [2,0,2,1,0,1,0,1,2]$, the algorithm starts with $E = [3,3,3]$. The algorithm iterates on $\sigma$ and starts with $val = \sigma[1] = 2$.

The loop that starts at step 5 and finishes at step 20, iterates all the value lower that 2 and compute the number of permutations with repetitions for each value with the following formula:
$$\frac{remaining!}{E[0]!\dots(E[j]-1)!\dots.E[n]!}$$

```
Algorithm 2. Compute_Permutation()
Input parameters:
   r : a rank
   [n] : an interval
Output parameters:
   σ : a permutation of n elements
Local parameters:
   E : List of element to insert in σ
Begin
1.  E = [m, m, … , m]
2.  r = 0
3.  remaining = n × m − 1
4.  For i = 0 to n × m − 2 do
5.     For j = 0 to n − 1 do
6.        If (E[j]>0) Then
7.        num = remaining!
8.        den = 1
9.        For k = 0 to n − 1 do
10.          If (k==j) and (E[k]>0) Then
11.             den = den × E[k − 1]!
12.          Else
13.             den = den × E[k]!
14.          End If
15.       End For
16.       Val = num × den
17.       If (Val < r) then
18.          r = r − Val
19.       Else
20.          Element = j
21.          break;
22.       End If
23.    End For
24.    σ[i] = Element
25.    E[Element] = E[Element] − 1
26.    remaining = remaining − 1
27. End For
28. j = 0
29. While E[j]==0 do
30.    j=j+1
31. End While
32. σ[i n × m − 1] = j
33. Return σ
34. End
```
Algorithm 2. Computation of $\sigma_r$

Because $val = \sigma[1] = 2$, the loop compute:
$$val = \frac{(8)!}{2!\,3!\,3!} = \frac{1032}{72} = 560$$
and the rank is currently $r = 560$
it computes next
$$val = \frac{(8)!}{3!\,2!\,3!} = \frac{1032}{72} = 560$$
and the rank is currently $r = 560 + 560 = 1120$

The first iteration is ended with $E = [3, 3, 2]$ modeling that only 2 number 2 remains available.
The second iteration of the first loop investigate $val = \sigma[2] = 0$ that do not update the current rank (there is no value lower that 0). The iteration number 3 investigate $val = \sigma[3] = 2$ and the final process gives the rank 1293.

The **algorithm 2** consists in computing the permutation $\sigma$ related to a rank $r$. The algorithm is composed of two loop: the first one (from step 4 to 27) iterates the $(n \times m)$ positions in the permutation $\sigma$ and the second loop (from step 9 to 29) investigates all the possible assignments of one job at the current position $i$. The job number that can be assigned to the current position are saved in $E = []$ with $E[j]$ is the number of job $j$ not yet included into permutation. So for one position $i$, let us note $remaining = n \times m − 1$ the number of positions in $\sigma$ with no job assigned, we have
$$val = \frac{remaining!}{E[0]!\dots(E[j]-1)!\dots.E[n]!}$$
that is the number if permutations with repetition with a $j$ number in position $i$ (step 16). The process is iterated by decreasing $r$ with $val$ until $val > r$ meaning that the current value of $j$ is the job to insert in position $i$.

Let us consider the rank 1293 related to the vector $[2, 0, 2, 1, 0, 1, 0, 1, 2]$. This vector is mapped into an oriented disjunctive graph (a solution) of makespan 188.

When $r = 1293$, the algorithm starts with $E = [3, 3, 3]$
$$\sigma = [\_,\_,\_,\_,\_,\_,\_,\_,\_]$$
At $i = 0$, the algorithm evaluated the number of permutation with repetition assuming that $\sigma$ starts with 0, starts with 1 and starts with 2: the evaluation for 0, 1 and 2 are investigated in the $j$ loop starting at step 5 and finishing at 23.

Assuming we have a 0 in first position, i.e.
$$\sigma = [0,\_,\_,\_,\_,\_,\_,\_,\_]$$
the number of permutation with repetition with a 0 in first position, 8 remaining positions with $E = [3, 3, 3]$ is
$$\frac{8!}{(E[0]-1)!\,E[1]!\,E[2]!} = \frac{8!}{2!\,3!\,3!} = \frac{40\,320}{72} = 560$$
Because $r = 1293 > 560$, we update r (step 21 in the algorithm) to $r = 1293 − 560 = 733$

Assuming we have a 1 in first position, i.e.
$$\sigma = [1,\_,\_,\_,\_,\_,\_,\_,\_]$$
the number of permutation with repetition with a 1 in first position, 8 remaining positions with $E = [3, 3, 3]$ is
$$\frac{8!}{(E[0])!\,(E[1]-1)!\,E[2]!} = \frac{8!}{3!\,2!\,3!} = \frac{40320}{72} = 560$$
Because $r = 733 > 560$, we update r (step 21 in the algorithm) to $r = 733 − 560 = 173$

Assuming we have a 2 in first position, i.e.
$$\sigma = [2,\_,\_,\_,\_,\_,\_,\_,\_]$$
the number of permutation with repetition with a 1 in first position, 8 remaining positions with $E = [3, 3, 3]$ is
$$\frac{8!}{(E[0])!\,(E[1])!\,(E[2]-1)!} = \frac{8!}{3!\,3!\,2!} = \frac{40320}{72} = 560$$
Because $560 > r = 173$, we conclude that the element to insert is 2 (step 23) and the step 8-26 loop end. The current permutation is now:
$$\sigma = [2,\_,\_,\_,\_,\_,\_,\_,\_] \text{ with } E = [3, 3, 2]$$

The process is iterated and the first loop ended with
$$\sigma = [2, 0, 2, 1, 0, 1, 0, 1,\_] \text{ and } E = [0, 0, 1]$$
And finally the last remaining job 2 is inserted into $\sigma$ giving the final permutation
$$\sigma = [2, 0, 2, 1, 0, 1, 0, 1, 2]$$

## 2.5. Indirect QAOA

The IQAOA algorithm uses $\vec{\beta}$ and $\vec{\gamma}$ weights parametrized a quantum state $|\psi(\vec{\beta},\vec{\gamma})\rangle$ that defines a solution rank $x$ with probability $|\langle x||\psi(\vec{\beta},\vec{\gamma})\rangle|^2$ and an expectation value estimated by sampling: $\langle\psi(\vec{\beta},\vec{\gamma})|C^p|\psi(\vec{\beta},\vec{\gamma})\rangle$. Each sampling gives a rank that can be evaluated by transforming the rank into the corresponding permutation and the permutation into a fully oriented disjunctive graph.

For a fixed $\vec{\beta},\vec{\gamma}$, the quantum computer is used to define the quantum state $|\varphi(\vec{\beta},\vec{\gamma})\rangle$ and the measure in the computational basis is achieved to get a qubit-string $x$ and evaluated $\langle\varphi(\vec{\beta},\vec{\gamma})|C^p|\varphi(\vec{\beta},\vec{\gamma})\rangle$.

The IQAOA has been successfully applied to the TSP (Bourreau et al., 2023), to the SAT (Fleury and Lacomme, 2023) and to the flow-shop (Fleury and Lacomme, 2024). IQAOA seeks to solve an optimization problem i.e. minimizing a function $m(x)$ acting on $n - bits$ strings modeling the rank of one solution. IQAOA is based on consecutive iterations of one Hamiltonian $H_P$ related to a driven Hamiltonian $H_D$.

The IQAOA efficiency relies on:
- A well-defined Hamiltonian $H_P$ that maps rank $r$;
- A well-defined mixer Hamiltonian $H_D$ that maps rank $r$;
- The availability of one iterative methods to investigate the parameters $\vec{\beta}$ and $\vec{\gamma}$ that minimize an objective function $C^p(\vec{\beta};\vec{\gamma})$ where $p$ is the number of shots required to estimate $C(\vec{\beta};\vec{\gamma})$.

(Bourreau et al., 2023) have introduced a binary representation of rank is

$$rank = \sum_{j=0}^{n} x_j . 2^j \text{ with } x_j \in \{0;1\} \quad (1)$$

with

$$H_P = \frac{1}{2}\sum_{j=0}^{n-1}(Id - Z_j).2^j \quad (2)$$

and for $H_D$ it is possible to use any parametrized quantum circuit that depends on $\theta$ commonly used in VQE i.e. single-qubit rotations and entangling drift operation acting on pairs of qubits (Amaro et al., 2022) (Moll et al., 2017).

In the rest of the paper we choose:

$$H_P(\gamma) = \frac{1}{2}\sum_{j=0}^{n}(Id - RZ_j(\gamma)).2^j$$

and depending on the instance

$$H_D^1(\beta) =. \left[\otimes_{j=0}^{n-1} RY_j(\beta)\right].\left[\otimes_{j=0}^{n-2} CX_{j,j+1}\right]$$

or

$$H_D^2(\beta) = \left[\otimes_{j=0}^{n-1} RX_j(\beta)\right].\left[\otimes_{j=0}^{n-2} CX_{j,j+1}\right].$$

or

$$H_D^3(\beta) =. \left[\otimes_{j=0}^{n-1} RY_j(\beta).RX_j(\beta)\right].\left[\otimes_{j=0}^{n-2} CX_{j,j+1}\right]$$

or

$$H_D^4(\beta) = \left[\otimes_{j=0}^{n-2} CX_{j,j+1}\right].\left[\otimes_{j=0}^{n-1} RY_j(\beta)\right]$$

which are inspired by the usual VQE proposals.

The transformation of a rank into a job-shop solution is achieved thanks to the algorithm 2 that transform a rank $r$ into a Bierwith's vector: $f:\{r\} \to \{\sigma\}$.

One permutation $\sigma$ is mapped into an acyclic disjunctive graph using the mapping function of (Bierwith, 1995) that gives the related makespan: $g:\{\sigma\} \to \mathbb{R}$

The full process is, for the $j^{th}$ measurement $|\psi(\vec{\beta},\vec{\gamma})\rangle_j$:

$$|\psi(\vec{\beta},\vec{\gamma})\rangle_j = r_j \mapsto f(r_j) = \sigma_j \mapsto g(\sigma_j)$$

To favor the metaheuristic convergence, at each optimization step $j$, we compute the average energy $\langle g^p\rangle_{\psi(\vec{\beta},\vec{\gamma})}$ that is expected to decrease, and the probability of the minimal energy that is expected to increase. Considering $p$ sampling we define:

$$C^p(\vec{\beta};\vec{\gamma}) = \xi.\langle g^p\rangle_{\psi(\vec{\beta},\vec{\gamma})} + \vartheta.M^p_{\psi(\vec{\beta},\vec{\gamma})}$$

Where $\xi$ and $\vartheta$ : are two number modeling the relative performance of $\langle g^p\rangle_{\psi(\vec{\beta},\vec{\gamma})}$ and $M^p_{\psi(\vec{\beta},\vec{\gamma})}$.

And

$$\langle g^p\rangle_{\psi(\vec{\beta},\vec{\gamma})} = \frac{1}{p}\sum_{j=1}^{p} g(\sigma_j)$$

$$M^p_{\psi(\vec{\beta},\vec{\gamma})} = \left[\min_{j=1..p} g(\sigma_j)\right] \times \left(p - \#\left\{k\Big|g(\sigma_k) = \min_{j=1..p} g(\sigma_j)\right\}\right)$$

All experiments were conducted utilizing Qiskit (IBM) on a simulator with $\xi = 100000$ and $\vartheta = 1$ to guarantee that $\min\langle g^p\rangle_{\psi(\vec{\beta},\vec{\gamma})} > \max M^p_{\psi(\vec{\beta},\vec{\gamma})}$.

The used metaheuristic is a genetic algorithm based on the **pygad** python library with 200 generations, 15 chromosomes per population, parent selection based on tournament, mutation probability of 0.70, and mutation percent genes of 25%. All the experiments use a quantum circuit of depth 2 for QAOA.

## 2.6. Resolution of a 3 jobs – 3 machines widget instance

The small scale instance 3× 3 is a job-shop with 3 jobs and 3 machines defining a total of 9 operations to schedule. The total number of permutations is 1680 but there is only 14 different makespans (table 1).

Table 1
Instance 3 jobs – 3 machines

| Job 1 | M1(21) | M0(5) | M2(10) |
|---|---|---|---|
| Job 2 | M0(11) | M1(15) | M2(16) |
| Job 3 | M2(39) | M0(100) | M1(42) |

Enumeration of all permutations show that the high quality solutions have relatively high probability (Fig. 1): the optimal solution 181 has a probability of about 55%.

Table 2
Initial distribution of permutations per makespan

| Makespan | Number of permutations | Probabilities |
|---|---|---|
| 181 | 928 | 55.24 |
| 194 | 81 | 4.82 |
| 207 | 116 | 6.90 |
| 212 | 225 | 13.39 |
| 217 | 75 | 4.46 |
| 222 | 84 | 5.00 |
| 223 | 30 | 1.79 |
| 228 | 15 | 0.89 |
| 232 | 12 | 0.71 |
| 233 | 56 | 3.33 |
| 243 | 33 | 1.96 |
| 248 | 11 | 0.65 |
| 249 | 9 | 0.54 |
| 259 | 5 | 0.30 |

The over-representation of high-quality solutions primarily stems from the fact that Bierwith vectors naturally define semi-active solutions on one hand, and on the other hand, they only model feasible solutions. However, such a significant value is a particular characteristic of the instance. We will later see that this figure will be much less significant for other instances.

```
0    :  [0, 0, 0, 1, 1, 1, 2, 2, 2]    249
1    :  [0, 0, 0, 1, 1, 2, 1, 2, 2]    217
2    :  [0, 0, 0, 1, 1, 2, 2, 1, 2]    217
3    :  [0, 0, 0, 1, 1, 2, 2, 2, 1]    217
….
1518 :  [2, 1, 2, 1, 0, 1, 2, 0, 0]    181
1519 :  [2, 1, 2, 1, 0, 2, 0, 0, 1]    181
1520 :  [2, 1, 2, 1, 0, 2, 0, 1, 0]    181
1521 :  [2, 1, 2, 1, 0, 2, 1, 0, 0]    181
1522 :  [2, 1, 2, 1, 1, 0, 0, 0, 2]    181
1523 :  [2, 1, 2, 1, 1, 0, 0, 2, 0]    181
1524 :  [2, 1, 2, 1, 1, 0, 2, 0, 0]    181
1525 :  [2, 1, 2, 1, 1, 2, 0, 0, 0]    217
1526 :  [2, 1, 2, 1, 2, 0, 0, 0, 1]    233
…
1677 :  [2, 2, 2, 1, 1, 0, 0, 1, 0]    232
1678 :  [2, 2, 2, 1, 1, 0, 1, 0, 0]    232
1679 :  [2, 2, 2, 1, 1, 1, 0, 0, 0]    232
```

For example, [2, 1, 2, 1, 0, 2, 0, 1, 0] of rank 1520 gives a solution of makespan 181.

Table 3
Final distribution of permutations per makespan (1000 samplings)

| Makespan | Number of permutations | Probabilities |
|---|---|---|
| 181 | 999 | 99.6 |
| 194 | 0 | 0.2 |
| 207 | 0 | 0.0 |
| 212 | 2 | 0.2 |
| 217 | 1 | 0.1 |
| 222 | 0 | 0.0 |
| 223 | 0 | 0.0 |
| 228 | 0 | 0.0 |
| 232 | 0 | 0.0 |
| 233 | 0 | 0.0 |
| 243 | 0 | 0.0 |
| 248 | 0 | 0.0 |
| 249 | 1 | 0.1 |
| 259 | 0 | 0.0 |

After the IQAOA execution with $H_D^1(\gamma)$, the sampling with 1000 shots gives 181 with 975 shots (table 3) meaning that about 97.4% of the probabilities is now on the optimal solution: the amplification is about 2 times. These results prove that IQAOA succeeds into transforming the amplitude of one uniform distribution into that of a target state.

*2.7. Resolution of a 4 jobs – 3 machines instance*

We consider a scale instance $4 \times 3$ is a job-shop with 4 jobs and 3 machines introduced in table 4.

Table 4
Instance 4 jobs – 3 machines

| | | | |
|---|---|---|---|
| Job 1 | M1(21) | M0(5) | M2(10) |
| Job 2 | M0(11) | M1(15) | M2(16) |
| Job 3 | M2(19) | M0(10) | M1(4) |
| Job 4 | M1(9) | M2(1) | M0(6) |

For this instance, we have 369599 permutations and the optimal solution value 59 which is related, for example, with:
$$\sigma = [0, 1, 0, 1, 2, 0, 1, 2, 2, 3, 3, 3]$$
A full enumerations of solution gives the representation of figure 6 proving that the probabilities to have a quality solution is rather low.

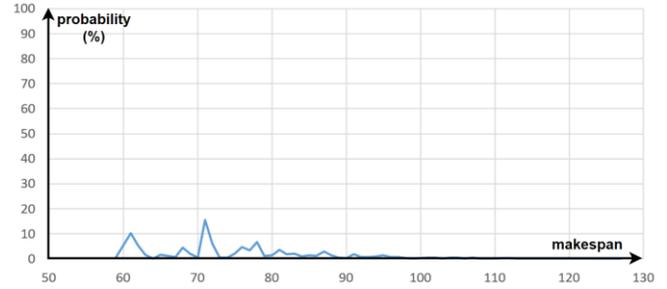
**Fig. 6** Initial distribution of makespan (4 jobs – 3 machines)

We have only a probability about 0.5 percent to find an optimal solution. The quartile is 68 meaning that we have only a probability of 25 percent to find a permutation mapped to a makespan lower than 68.

Table 4
4-jobs 3-machines Job-Shop instances

| Makespan | Number of permutations | Probabilities |
|---|---|---|
| 59 | 1952 | 0.53 |
| 61 | 37999 | 10.28 |
| 62 | 19582 | 5.30 |
| 63 | 5904 | 1.60 |
| 64 | 694 | 0.19 |
| 65 | 6064 | 1.64 |
| 66 | 4776 | 1.29 |
| 67 | 3067 | 0.83 |
| 68 | 16891 | 4.57 |
| 69 | 8062 | 2.18 |
| 70 | 2072 | 0.56 |
| 71 | 57913 | 15.67 |
| 72 | 22711 | 6.14 |
| 73 | 1872 | 0.51 |
| 74 | 1879 | 0.51 |
| 75 | 8015 | 2.17 |
| 76 | 17861 | 4.83 |
| 77 | 12806 | 3.46 |
| 78 | 25151 | 6.80 |
| 79 | 4292 | 1.16 |
| 80 | 5400 | 1.46 |
| 81 | 13423 | 3.63 |
| 82 | 6972 | 1.89 |
| 83 | 7800 | 2.11 |
| 84 | 3392 | 0.92 |
| 85 | 5506 | 1.49 |
| 86 | 4357 | 1.18 |
| … | … | … |
| 121 | 70 | 0.02 |
| 122 | 64 | 0.02 |
| 123 | 80 | 0.02 |
| 126 | 6 | 0.00 |
| 127 | 26 | 0.01 |

After optimization using $H_D^4(\gamma)$ and 200 generations of the GA, we obtain a probability of 99% on permutation of makespan 59 as stressed on table 5.

Table 5
4-jobs 3-machines Job-Shop instances: final distribution

| Makespan | Number of shots | Probabilities |
|---|---|---|
| 59 | 997 | 99.7 |
| 62 | 2 | 0.2 |
| 91 | 1 | 0.1 |

## 2.8. Resolution of a 5 jobs – 2 machines instance

We consider a scale instance 5 × 2 is a job-shop with 4 jobs and 3 machines introduced in table 6.

Table 6
Instance 5 jobs – 3 machines

| Job 1 | M1(2) | M0(4) |
|---|---|---|
| Job 2 | M1(1) | M0(5) |
| Job 3 | M0(6) | M1(5) |
| Job 4 | M0(4) | M1(3) |
| Job 5 | M0(3) | M1(7) |

The IQAOA using 200 generations and $H_D^2(\gamma)$ gives a final distribution (Table 7) where 99% of the distribution is on the optimal makespan: one permutation that is mapped into a makespan of 22 is, for example: $\sigma = [1, 4, 3, 0, 2, 4, 0, 3, 1, 2]$

Table 7
5-jobs 2-machines Job-Shop instances initial and final distribution

| Makespan | Probabilities of the initial distribution | Probabilities of the final distribution |
|---|---|---|
| 22 | 16.11 | 99.60 |
| 23 | 5.71 | 0.10 |
| 24 | 8.51 | 0.20 |
| 25 | 12.96 | 0.10 |
| 26 | 5.32 | |
| 27 | 8.44 | |
| 28 | 9.46 | |
| 29 | 7.54 | |
| 30 | 6.38 | |
| 31 | 4.57 | |
| 32 | 3.98 | |
| 33 | 3.11 | |
| 34 | 3.42 | |
| 35 | 1.38 | |
| 36 | 1.02 | |
| 37 | 1.15 | |
| 38 | 0.33 | |
| 39 | 0.33 | |
| 40 | 0.28 | |

Let us note that the convergence to a distribution concentrated on the optimal solution required only 142 fitness evaluations and the parameters $\vec{\beta}$ and $\vec{\gamma}$ converge in a short number of iterations to the final state (table 8).

Table 8
Convergence of $\vec{\beta}$ and $\vec{\gamma}$

| $\vec{\beta}$ | | $\vec{\gamma}$ | | Makespan |
|---|---|---|---|---|
| -0.57682 | -0.36004 | -2.49850 | 2.74524 | 29039719 |
| -2.02314 | 2.69230 | -1.38641 | -2.97782 | 27476740 |
| 2.16826 | 1.69198 | -2.85352 | 0.55557 | 26939718 |
| -2.60440 | -0.12792 | -1.13131 | -0.06045 | 25489476 |
| -3.19286 | 2.58213 | -0.31559 | -1.45345 | 25401233 |
| -2.90900 | 3.49420 | 0.63200 | -1.10877 | 25275991 |
| -3.46656 | 3.00122 | -0.37044 | -1.45345 | 25113881 |
| -3.10826 | 2.76945 | 0.36952 | -0.02074 | 24275661 |
| -3.00476 | 2.83087 | 1.04985 | -1.07214 | 23488528 |
| -2.90900 | 3.00122 | -0.25591 | -0.91068 | 23363440 |
| -3.25337 | 3.06065 | -0.23755 | -1.55068 | 22225066 |
| -3.25337 | 3.06065 | 0.57168 | -1.80382 | 22150066 |
| -3.09423 | 3.16680 | 1.34698 | -1.82660 | 22000000 |

## 2.9. Resolution of a 3 jobs – 4 machines instance

We consider a scale instance 3 × 4 that is a job-shop with 3 jobs and 4 machines introduced in table 9. This is a small scale instance with 12 operations to schedule and 34 650 permutations.

Table 9
Instance 3 jobs – 4 machines

| Job 1 | M1(2) | M0(4) | M2(8) | M3(2) |
|---|---|---|---|---|
| Job 2 | M3(10) | M0(11) | M1(1) | M2(5) |
| Job 3 | M2(9) | M3(5) | M0(3) | M1(1) |

The IQAOA using 200 generations and $H_D^3(\gamma)$ gives a final distribution (Table 10) where 61% of the distribution is on the optimal makespan.

Table 10
Results for the 3 jobs – 4 machines

| Makespan | Probabilities of the final distribution | Probabilities of the initial distribution |
|---|---|---|
| 27 | 14.47 | 61.20 |
| 28 | 2.08 | 2.40 |
| 31 | 2.45 | 5.00 |
| 32 | 3.77 | 1.90 |
| 34 | 0.17 | 0.00 |
| 35 | 5.54 | 5.80 |
| 37 | 8.49 | 7.90 |
| 38 | 23.06 | 3.80 |
| 39 | 1.82 | 0.20 |
| 41 | 19.40 | 7.40 |
| 42 | 0.10 | 0.00 |
| 43 | 0.20 | 0.00 |
| 44 | 1.48 | 0.20 |
| 45 | 2.19 | 0.10 |
| 46 | 3.52 | 0.80 |
| 47 | 0.17 | 0.00 |
| 48 | 0.45 | 0.10 |
| 49 | 0.28 | 0.10 |
| 50 | 0.10 | 0.00 |
| 51 | 3.63 | 1.40 |
| 52 | 4.07 | 1.10 |
| 53 | 0.20 | 0.00 |
| 54 | 0.15 | 0.00 |
| 55 | 1.60 | 0.30 |
| 56 | 0.18 | 0.10 |
| 57 | 0.13 | 0.10 |
| 58 | 0.22 | 0.10 |
| 59 | 0.06 | 0.00 |
| 61 | 0.01 | 0.00 |

The significant difference between the two distribution is highlighted on figure 7.

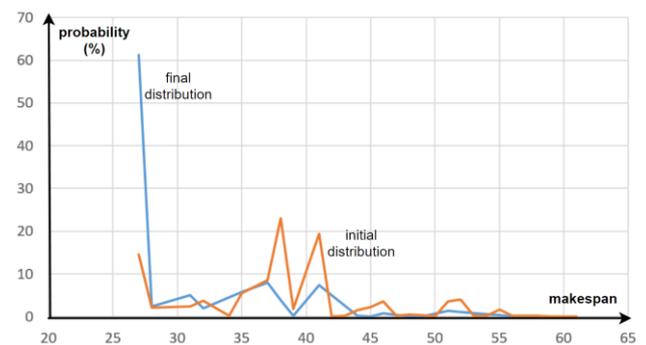

**Fig. 7.** Final and initial distribution of makespan

## 2.10. Resolution of a 4 jobs – 4 machines instance

We consider a large scale instance 4 × 4 that is a job-shop with 4 jobs and 4 machines introduced in table 11. This is a large scale instance with 16 operations to schedule and 63 063 000 of permutations and there are 113 different values for the makespan.

Table 11
Instance 4 jobs – 4 machines

| Job 1 | M1(21) | M3(10) | M0(5)  | M2(10) |
| Job 2 | M0(11) | M1(15) | M3(25) | M2(16) |
| Job 3 | M2(19) | M0(10) | M1(4)  | M3(2)  |
| Job 4 | M3(80) | M1(9)  | M2(1)  | M0(6)  |

The full enumeration of permutations required about 3 hours on a classical computer and this enumeration shows that only 7.7% of the distribution is on makespan 131 and the quartile is 145 (the probability to have a makespan lower that 145 is about 25%). The best solution makespan is 131 and this makespan is related, for example, with

$$\sigma = [0, 1, 1, 2, 2, 2, 3, 0, 0, 0, 1, 2, 3, 3, 1, 3]$$

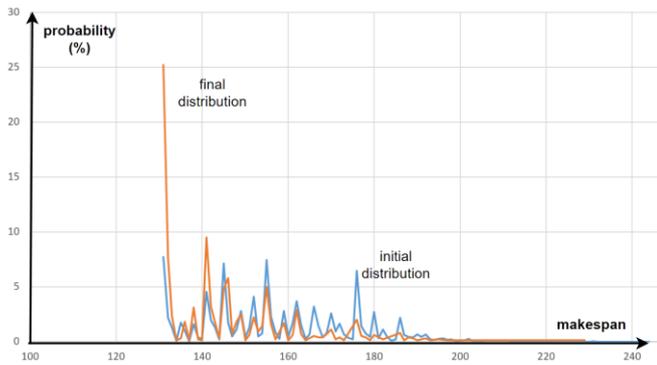

**Fig. 8.** Final and initial distribution of makespan for the 4 jobs-4 machines instance

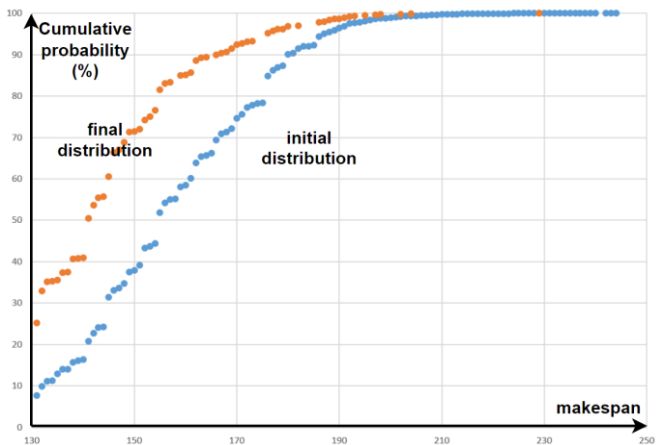

**Fig. 9.** Final and initial cumulative distributions

At the end of the optimization using $H_D^2(\gamma)$, 25.2% of the distribution is on makespan 131 and the quartile is 131 (figure 8) proving the efficiency of the optimization. The cumulative distributions of figure 9 emphasis the impact of the optimization on the distribution.

## 3. Conclusion

We have evaluated the effectiveness of IQAOA in solving the Job-Shop problem, and the findings indicate that small-scale instances (the largest being a 4 jobs-4 machines instance) can be effectively tackled using this method. This approach proves applicable to all problems where the indirect representation simplifies to computing a rank. However, it necessitates the definition of a finely tuned mixer to ensure efficiency in practically resolving Job-Shop instances.